\documentclass[]{spie}

\usepackage{amsmath,amsfonts,amssymb}
\usepackage{graphicx}
\usepackage[colorlinks=true, allcolors=blue]{hyperref}
\usepackage{graphicx}
\usepackage{subcaption}
\usepackage{dirtytalk}  

\title{Towards Image Synthesis with Photon Counting Stellar Intensity Interferometry}

\author[a]{Alessia Spolon}
\author[a]{Michele Fiori}
\author[a]{Luca Zampieri}
\author[b]{Marco Landoni}
\affil[a]{INAF - Osservatorio Astronomico di Padova, Vicolo dell’Osservatorio 5, 35122 Padova, Italy}
\affil[b]{INAF - Osservatorio Astronomico di Brera, Via Brera, 28, 20121 Milano, Italy}

\authorinfo{Further author information: (Send correspondence to Alessia Spolon)\\A.S.: E-mail: alessia.spolon@inaf.it \\  M.F.: E-mail: michele.fiori@inaf.it \\ L.Z.: E-mail: luca.zampieri@inaf.it \\
M.L.: E-mail: marco.landoni@inaf.it}

\begin{document} 
\maketitle

\begin{abstract}

Stellar intensity interferometry (SII) is based on the correlation of the light intensity fluctuations of a star detected at two or more telescopes, with no need to combine the collected photons directly. A measurement of the correlation in full "photon-counting mode" was experimented with fast photon counters in Italy (2016-2020) and is currently being adapted to the ASTRI Mini-Array. Performing image synthesis with "photon-counting" SII requires a series of preparatory activities that involve the optimization of the pipelines for the treatment of time series acquired at extremely high photon rates, the development of efficient and innovative algorithms for the cross-correlation of the arrival times in large time series and the development of a preliminary version of a dedicated pipeline for the synthesis of images starting from interferometric data. Here we present the project and the present status of the activities.

\end{abstract}
 

\section{INTRODUCTION}
\label{sec:intro}  

\subsection{Stellar Intensity Interferometry (SII)}

\begin{figure} [ht!]
   \begin{center}
   \includegraphics[width=0.7\textwidth]{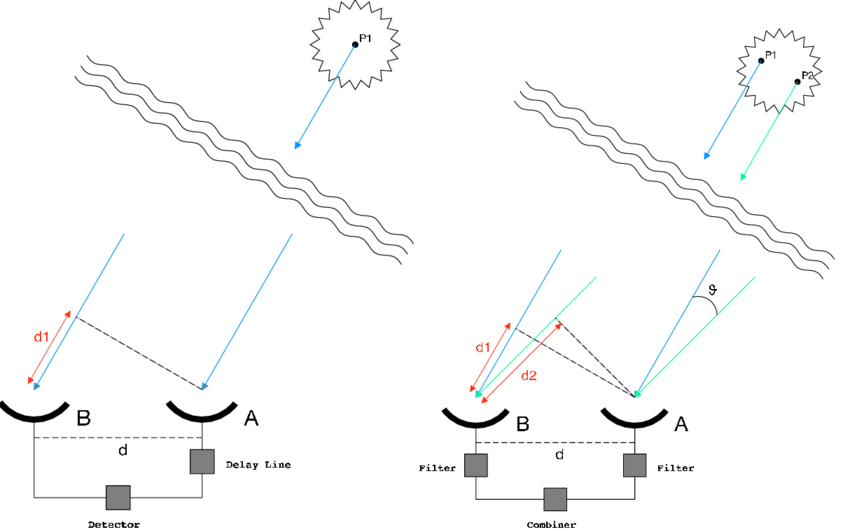}
   \end{center}
   \caption[example] 
   { \label{fig:foellmi} 
Working principle of an intensity interferometer compared to that of a Michelson interferometer. \textit{On the left} (Michelson interferometer): a first-order spatial coherence measurement is made, associated with the statistical average of single point source. Delay lines are used to cancel out the path difference $d_{1}$ and ensure temporal coherence. \textit{On the right} (intensity interferometer): a second-order spatial coherence measurement is made, associated with the statistical average of the correlations between pairs of point sources on the surface of a star (\textit{$P_{1}$}, \textit{$P_{2}$}). $\theta$ is the angular separation of the two points \textit{$P_{1}$} and \textit{$P_{2}$} on the star; $d$ is the distance between the two telescopes. Figure from [\citenum{Foellmi09}]. }
\end{figure}

\noindent Nowadays, we have the capability to image bright stars in the visible light waveband at very high angular resolution using a technique known as Stellar Intensity Interferometry (SII), which is based on the measurement of the second order coherence of light [\citenum{glauber63}]. Angular resolutions below 100 micro-arcsecond $(\mu$as) are achievable with this technique, using large collecting area telescopes separated by hundreds to thousands of meters baselines. At this level of resolution it turns out to be possible to reveal details on the surface and of the environment surrounding bright stars on the sky, that typically have angular diameters of 1-10 milli-arcsecond (mas) [\citenum{kieda19}]. \\
Stellar intensity interferometry measures the correlation of the light intensity fluctuations of a star detected at two or more telescopes (Fig. \ref{fig:foellmi}, right panel;  [\citenum{Foellmi09}]), unlike  ordinary amplitude interferometry which measures the fringes generated by the direct interference of the telescopes light beams (e.g. the Michelson interferometer, see Figure \ref{fig:foellmi}: left panel; [\citenum{Foellmi09}]). \\

\begin{figure} [ht!]
   \begin{center}
   \includegraphics[width=7cm, height=5cm]{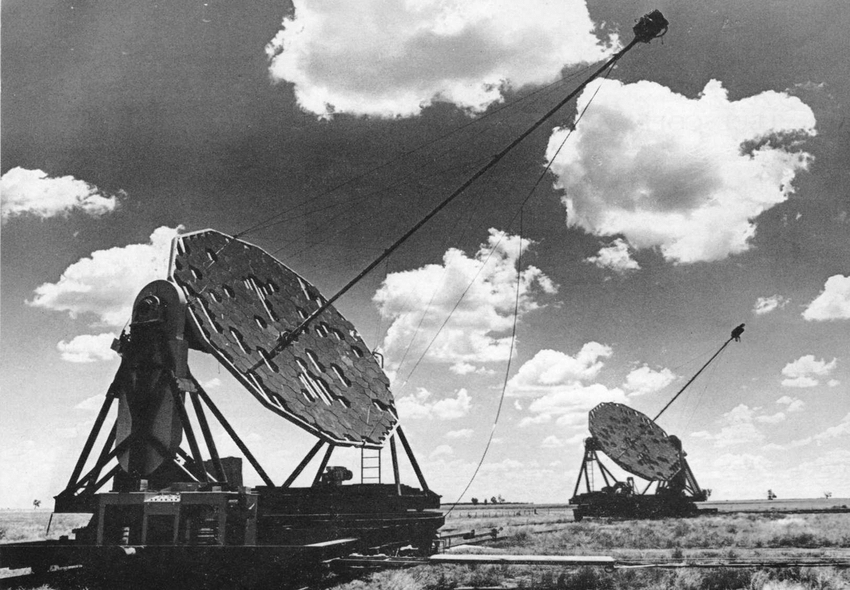}
   \end{center} 
   \caption[example] 
   { \label{fig:narrabi}
   Light collectors used in the Narrabri stellar intensity interferometer. Image from Sky \& Telescope, vol. 28, pp. 2-7, 1964. }
   \end{figure}

\noindent SII was pioneered by Robert Hanbury Brown and Richard Q. Twiss between the 50s and the 70s [\citenum{Brown56, Brown57, Brown58, Brown74}]. 
They built the Narrabri Stellar Intensity Interferometer using twin 6.5 m diameter telescopes movable along a circular track at Narrabri, New South Wales, Australia (see Figure \ref{fig:narrabi}), and performed the first direct astronomical measurements of stellar radii via SII. After the successful Narrabri experiment, SII was shelved for about 40 years. \\

\noindent The possibility to operate simultaneously an array of large area telescopes and to connect them electronically, with no need to directly combine the photons they detect, has recently renewed interest for SII as a tool for performing imaging observations in the optical band using a detection method similar to long-baseline radio interferometric arrays [\citenum{Bohec06, Dravins13}]. 
This possibility is offered by the sparsely distributed arrays of Imaging Air Cherenkov Telescopes (IACTs), such as the ASTRI Mini-Array, which have suitable optical properties, sufficiently large mirror areas and very fast electronics.
IACTs are telescopes designed for gamma-ray observations (they look for Cherenkov flashes due to the interaction of gamma-ray photons with the atmosphere) and cannot observe during the nights around the full moon, allowing these telescopes to be used in SII mode for several nights each month. 
An advantage of SII is that the precision of the measurement of photon arrival times can be of the order of one ns at each telescope, over baselines extending to km distances. This accuracy corresponds to tenths of meter light-travel distance, and thus any instrumental or atmospheric delay smaller than a fraction of one meter can be tolerated. 
New implementations of SII technology to Astronomy have been recently pursued by several groups, performing pilot experiments or observations with 1-3 meter class optical telescopes 
[\citenum{Zampieri_2016, Zampieri21, Guerin17}], or using current  IACTs systems.
The VERITAS, MAGIC and HESS telescopes have already shown the enormous potentiality of performing SII measurements using arrays of Cherenkov telescopes [\citenum{abe24, Abeysekara_2020, Zmija_2023}]. 

\subsection{ASTRI Mini-Array}
\begin{figure}[htbp]
  \centering
  \begin{subfigure}{0.407\textwidth}
    \centering
    \includegraphics[width=\textwidth]{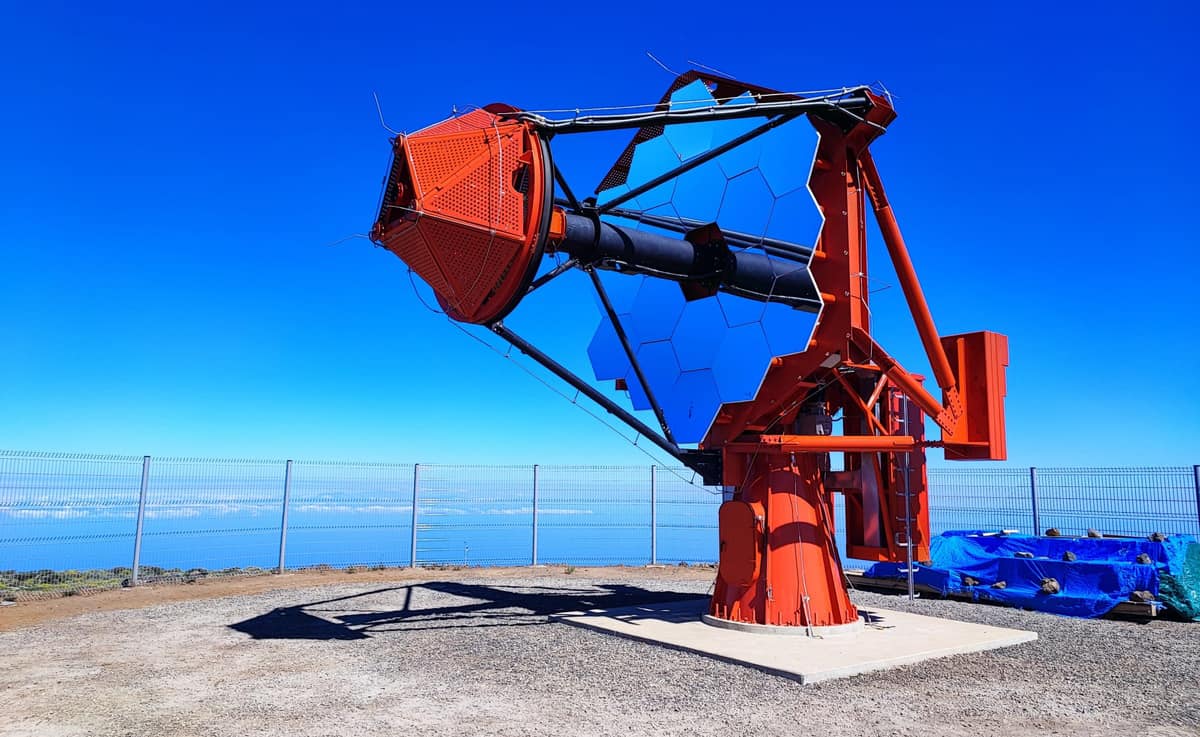}
  \end{subfigure} 
  \hspace{0.1\textwidth}
  \begin{subfigure}{0.44\textwidth}
    \centering
    \includegraphics[width=\textwidth]{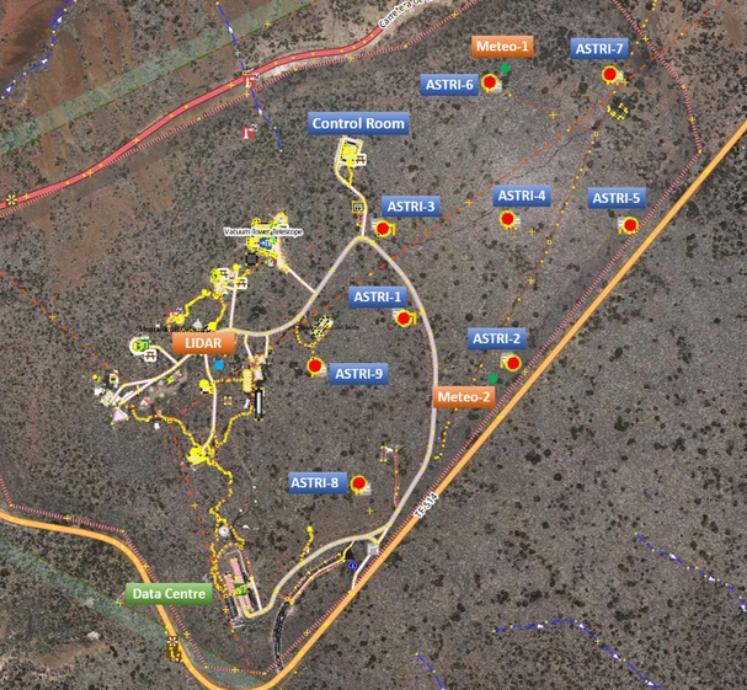}
  \end{subfigure}
  \vspace{0.8\baselineskip} 
  \smallskip
  \caption{\label{fig:astri} \textit{Left:} The first of the 9 ASTRI telescopes mounted on site. \textit{Right:} Final layout of the ASTRI Mini-Array at the site of El Teide in Tenerife, Spain. From [\citenum{Scuderi22}].}
\end{figure}

Since the beginning of 2019 also the INAF ASTRI (Astrophysics with Italian Replicating Technology Mirrors) Collaboration recognizes the scientific value of SII and supports the development of a SII observing mode for the Mini-Array. The ASTRI project was approved in 2010 to support the development of technologies within the Cherenkov Telescope Array project. In this framework INAF will build an independent Mini-Array of 9 Cherenkov 4m-class telescopes in Schwarzschild-Couder optical configuration in Tenerife (Spain) (see Figure \ref{fig:astri}; [\citenum{Scuderi22}]). The ASTRI Mini-Array will offer an ideal SII imaging installation thanks to the capabilities offered by its 9 telescopes, that provide 36 simultaneous baselines over distances between 100 m and 700 m [\citenum{Vercellone_2022}].

\noindent The ASTRI Mini-Array will be equipped with a SII Instrument (SI$^3$) for performing images of bright Galactic stars with sub-mas angular resolution [\citenum{Zampieri22}]. Measuring the angular shape of a selected number of stars (including main sequence stars) with a resolution of $\sim$ 100 $\mu$as will allow for the determination of their oblateness and enable direct measurements of the stellar rotation.
Imaging with this resolution can also allow the detection of dark/bright spots or other surface features [\citenum{Nu_ez_2012}]. 
Furthermore, observing stars with circumstellar discs/eruptions will reveal details of the disc structure, density gradients, and scale height, and will show how these systems evolve and dynamically interact [\citenum{Kloppenborg10, Fiori21}]. Concerning this, the operation of the ASTRI Mini-array in SII mode will establish a remarkable heritage of sub-mas images of the brightest nearby stars and their environments. 

\noindent For SII observations the optimal targets are stars with high brightness temperature, that have both a significant photon flux and structures small enough to produce coherence over long baselines. Therefore, O-thorough-G type stars of adequate brightness are all suitable and potential targets, which makes the B band (between 420 nm and 500 nm) the appropriate working wavelength window.

\section{METHODS}

\subsection{Measurements of the degree of coherence}

The main observable for SII is the second-order (discrete) degree of coherence $g^{(2)}(\tau,d)$  of a star, that measures the degree of correlation of its light and depends on the telescopes/sub-apertures separation $d$ and the relative delay $\tau$ between them. 
We can calculate $g^{(2)}(\tau,d)$ in photon counting and in post-processing using the expression [\citenum{Naletto16, Zampieri_2016}]: 
\begin{equation}
g^{(2)}(\tau,d)=\frac{N_{XY} N} {N_{X} N_{Y}}  \label{eq:g2}
\end{equation}
where $N_{X}$ and $N_{Y}$ are the number of photons detected on the sub-apertures X and Y (of the same telescope or of two different telescopes) in a time interval $T$, $N_{XY}$ is the number of simultaneous detections (coincidences) in both sub-apertures in a small time bin \textit{dt}, and $N = T/dt$ is the total number of bins in time $T$. \\
The measurement of $g^{(2)}(\tau,d)$ can be entirely performed in post-processing using a dedicated software package (written in \texttt{Linux} bash shell, \texttt{Fortran}, \texttt{Python}), that computes in an efficient way the number of coincidences in the photon arrival times at the different detectors or at the two telescopes. \\
The search for coincidences $N_{XY}$ is performed after binning the event lists related to two different detectors using a certain time bin. Two photons are considered coincident only if they fall in the same bin. The algorithm is optimized in such a way that it registers only the time bins in which the detection of a photon actually occurs, discarding all others and thus reducing the computation time [\citenum{Zampieri21}]. \\

\subsubsection*{The Asiago Stellar Intensity Interferometry experiment}

To demonstrate the feasibility to perform photon-counting SII measurements, we used two of the most advanced photon counters available for astronomy: Aqueye+ [\citenum{Barbieri09, Naletto13, Zampieri15}]  and Iqueye [\citenum{Naletto09}]. 
The observing facilities are the 1.22 m Galileo telescope (T122), located at the Pennar station, and the 1.82 m Copernicus telescope, located at Cima Ekar. The telescopes are about 3.9 km apart from each other with a predominant East-West orientation [\citenum{Naletto16}]. 
With Aqueye+ and Iqueye installed, the two telescopes are well suited to realize a photon counting km-baseline intensity interferometry. 
Iqueye is connected to the T122 through the Iqueye Fiber Interface (IFI, [\citenum{Zampieri19}]), while Aqueye+ is directly mounted at the focus of the T182 telescope. 
These instruments share an identical instrumental framework, featuring the same detectors and time acquisition systems, thereby ensuring a relative time resolution better than 100 ps. 
For a comprehensive overview of the instruments, see [\citenum{Naletto09}]. 

\noindent With the Asiago SII experiment, by analyzing the coincidences in post-processing, we could not only perform the first measurements of this kind, but we could also validate the feasibility of such measurements on a kilometer-long baseline.

\subsection{Accelerating the Computation}

\noindent In order to perform image synthesis with photon-counting SII using the ASTRI Mini-Array, several preparatory activities are necessary. Firstly, optimizing the existing pipelines to handle time series acquired at extremely high count rates across the entire array is essential. Secondly, developing efficient and innovative algorithms for cross-correlating arrival times in large time series is a key activity. Finally, a preliminary version of a dedicated pipeline for synthesizing images from interferometric data and data related to instrumental simulations needs to be developed. \\
These activities involve not only scaling the computation of photon coincidences to handle extremely high count rates across the entire array (in the Big Data regime) but also necessitates the development, acceleration, and optimization of cross-correlation algorithms applied to large time series. 

\begin{figure} [ht!]
   \begin{center}
   \includegraphics[width=0.6\textwidth]{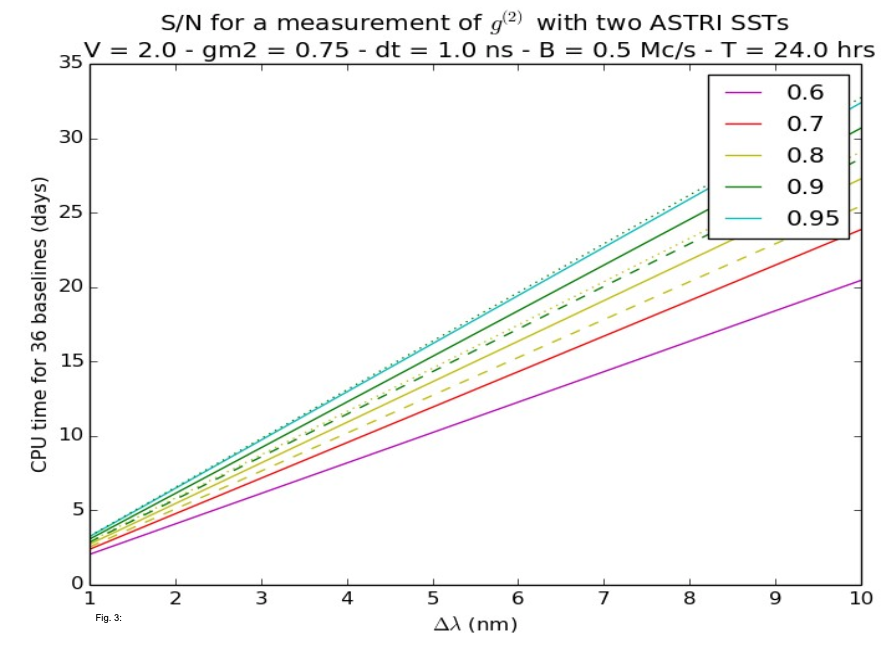}
   \end{center}
   \caption[example] 
   { \label{fig:cpuTime} Estimated computational time as a function of the bandpass of the filter used in the observations and of the optical throughput. With a 3 nm filter with 60\% of transmission (purple line), we estimate about 5 days of CPU time (for 36 baselines).     }
   \end{figure}

\noindent The idea is to use High-Performance Computing (HPC) systems, to efficiently perform these resource-intensive calculations. To analyze 1 hour of data, considering all the 36 baselines, requires 10$^{4}$ hours of central processing unit (CPU) time. This translates to 5 hours of processing time when utilizing 2000 cores. Therefore, to process the equivalent of 24 hours of real data (one entire observing run around full moon), it would take 5 days (5 $\times $ 24 hours). Figure \ref{fig:cpuTime} shows the estimated computation time based on the properties of the filter used in the observations and the and the optical throughput. \\
Our first goal is speeding up the computation optimizing the algorithm, employing multiple CPUs ($\sim2000$ CPU cores) in order to process 1 hour of data within $\sim$60 minutes. To do so, we requested for resources on a High-Performance Computing system (100 hours of SII data to optimize the code). However, by utilizing GPUs, we plan to substantially decrease the processing time by a factor of 20. Therefore we also required additional machine time to work on accelerating the algorithm using GPUs. 
To accelerate and parallelize algorithms we will adopt Compute Unified Device Architecture (CUDA), developed by NVIDIA [\citenum{cuda}]. CUDA is a platform allowing developers to utilize NVIDIA graphical processing units (GPUs) for general-purpose computing tasks beyond graphics rendering. This enables significant speedups for various tasks like scientific simulations, deep learning, and image processing by leveraging the computational power of GPUs. \\

\subsection{ASTRI SII Simulations}
To assess the capabilities of SI$^3$ in performing II measurements, we developed a dedicated \texttt{Python} software, that simulates the entire observation process, starting from the acquisition of the data up to post-processing. 
The software is designed to be modular and can easily be extended by integrating different functionalities, like new models for targets.  
In the simulation procedure we take into account: the technical specifications of the array that will be used for the observations (e.g. total number and size of the telescopes, efficiency of optical systems, optical filter central wavelength and bandwidth), the properties of the targets (e.g. spectrum, intrinsic brightness, expected angular size) and all the sources of noise that can be present (e.g. total brightness of the sky with and without moon, dark counts of the detectors).

\begin{figure} 
  \centering
  \begin{subfigure}{0.49\textwidth}
    \centering
    \includegraphics[width=\textwidth]{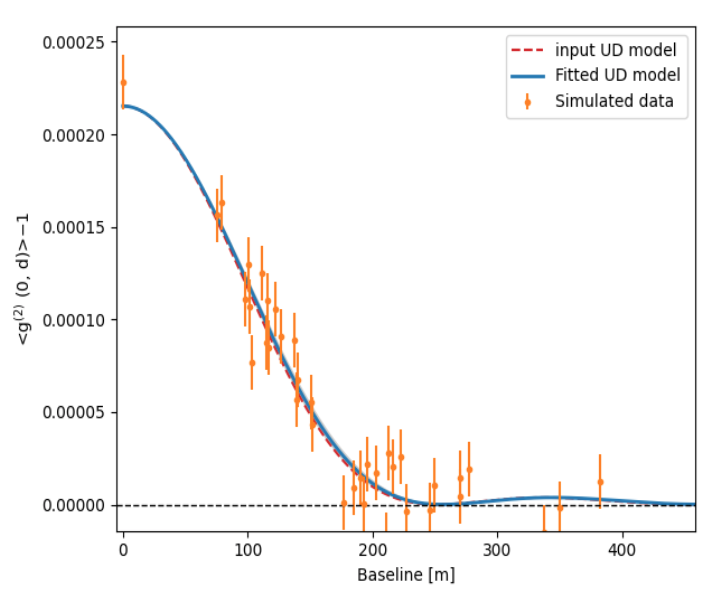}
    \label{fig:V_1.83}
  \end{subfigure} 
  \begin{subfigure}{0.48\textwidth}
    \centering
    \includegraphics[width=\textwidth]{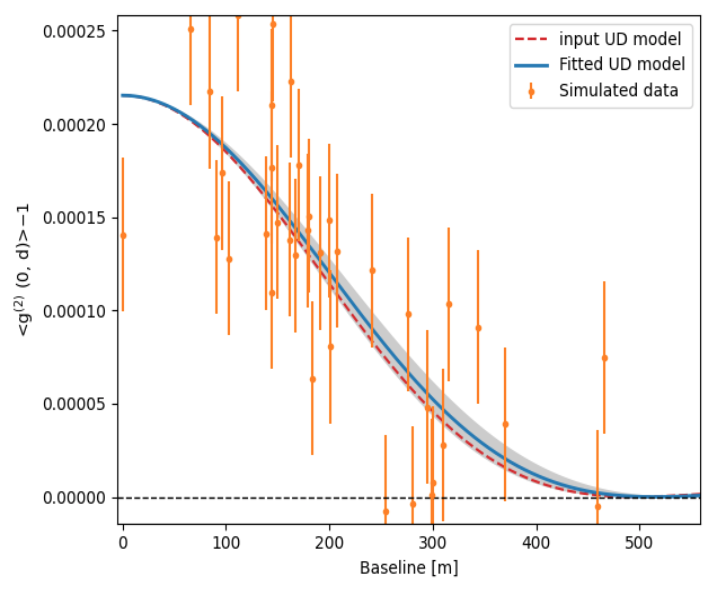}
    \label{fig:V_4.48}
  \end{subfigure}
  \smallskip
  \caption{\label{fig:astri_simul} Simulated $g^{(2)}$ measurements of two different stars with an uniform disk. In the left panel we show the fit of a simulation of a O7 type star with a magnitude V$=1.83$. With a total observing time of 10 hours we obtain a fitted diameter of $(0.432\pm0.009)$ mas ($\sim$2\% accuracy). In the right panel we show the fit of a simulation of a B3 type star with a magnitude V$=4.48$. With a total observing time of 30 hours in dark time we obtain a fitted diameter of $(0.215\pm0.012)$ mas ($\sim$5\% accuracy).}
\end{figure}

\noindent To understand the real performance of SI$^3$ and compare it withthat obtained with similar facilities (as VERITAS [\citenum{Abeysekara_2020}], MAGIC [\citenum{abe24}], HESS [\citenum{Zmija_2023}]), we carried out 1D simulations using a uniform disk model for the brightness distribution of all the stars observed from El Teide. The uniform disk is the simplest model that can be used to estimate with a good approximation the angular size of a star and is also a good indicator for assessing the capabilities of a SII system. In fact, for two-telescopes SII system, the Signal-to-Noise Ratio (SNR) of the measurement of the angular diameter of a star can be directly estimated from the theoretical SNR of a SII measurement [\citenum{Fiori22}].
Considering the 36 available baselines, we obtain angular diameter measurements with high SNR ($\sim50-100$ or $1-2$\% accuracy level) with only few hours of observations for bright targets (V$\leq2$). This is shown in the left panel of Fig. \ref{fig:astri_simul}, where we have simulated the observation of a O7 type star with a magnitude V$=1.83$ (average count rate $\sim30$ Mct/s) and an estimated angular diameter of 0.44 mas. With a total observing time of just 10 hours we can obtain a fitted diameter of $(0.432\pm0.009)$ mas ($\sim$2\% accuracy). In case of fainter stars (V$\leq5$), thanks to the possibility of sampling large part of the Visibility curve allows us to reach good results with a reasonable observing time. This is shown in the right panel of Fig. \ref{fig:astri_simul} where we simulated the observation of a B3 type star with a magnitude V$=4.48$ (average count rate $\sim4$ Mct/s) and an estimated angular diameter of 0.22 mas. With a total observing time of 30 hours in dark time we obtain a fitted diameter of $(0.215\pm0.012)$ mas ($\sim$5\% accuracy).

\begin{figure} 
  \centering
  \begin{subfigure}{0.46\textwidth}
    \centering
    \includegraphics[width=\textwidth]{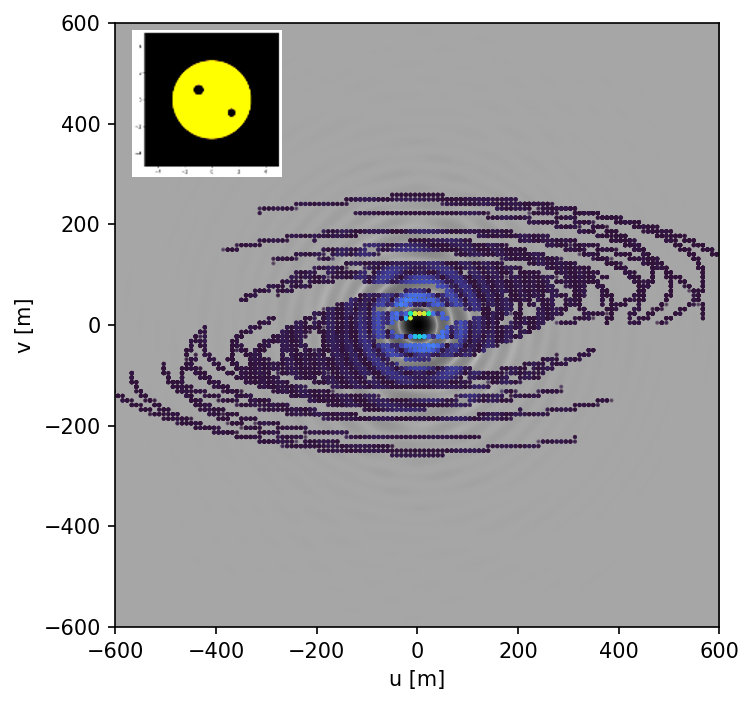}
  \end{subfigure} 
  \begin{subfigure}{0.525\textwidth}
    \centering
    \includegraphics[width=\textwidth]{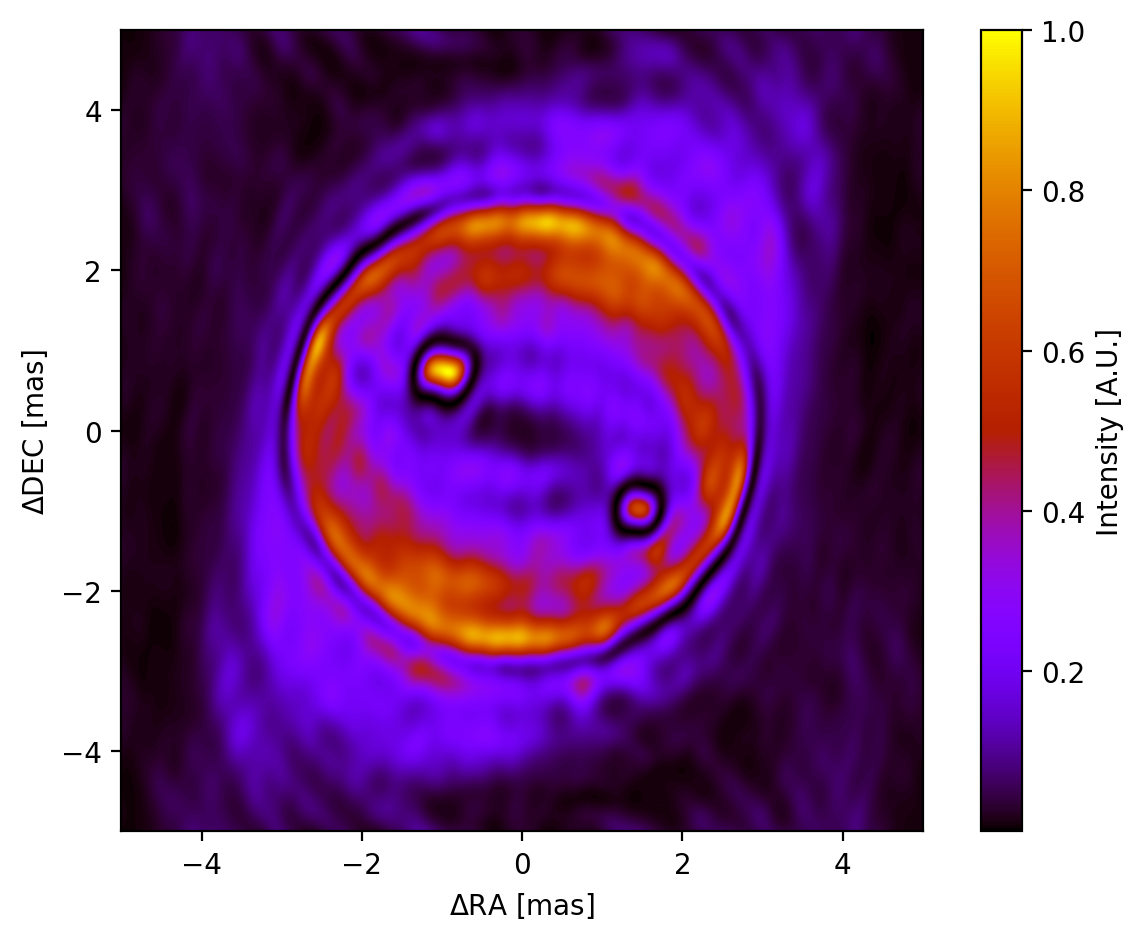}
  \end{subfigure}
  \smallskip
  \caption{\label{fig:astri_imaging} Simulations of an ASTRI SI$^3$ observation of an A-type star with some dark spots on the surface. In the left panel we show the 2D Fourier transform corresponding to the brightness distribution of the star, with overplotted the coverage attainable in two consecutive nights. In the right panel we show a possible reconstructed image considering that coverage.}
\end{figure}

\noindent The ultimate goal of the SI$^3$ project is to demonstrate the feasibility of direct SII imaging. This will be crucial in light of future implementations of SII on the Cherenkov Telescope Array [\citenum{Kieda19_sii3}]. Thus, the last activity related to this project will be the development of a dedicated pipeline for the synthesis of images using II data.
As a starting point, we began to produce a series of simulations aiming at understanding the best observational strategy and image reconstruction techniques.
In Fig. \ref{fig:astri_imaging} we show a simulation of a star with  dark spots (inset plot in left panel), from which we have calculated the complex Visibility using the van Cittert-Zernicke theorem [\citenum{Mandel95}]. In the panel on the left we show the 2D Fourier transform corresponding to the brightness distribution of the star, with overplotted the coverage attainable in two consecutive nights of observation using all the 36 available baselines. In the right panel we show the possible reconstructed image considering that coverage.
Unlike the fits of the 1D visibility through analytical models (e.g. Uniform Disks as shown before), in the U-V plane it is
clear how crucial it is to sample the Visibility away from the central peak. Therefore, the best scenario for the ASTRI Mini-array will be to try to observe stars with angular diameters around 1-3 mas in which substructures (like dark spots, disk of debris, transiting exoplanets, etc.) may exist.
Of course, image reconstruction as shown in this example will only be possible by adopting appropriate algorithms for phase reconstruction, since II is a second-order effect for which the information about the phase is lost at the moment of the measurements [\citenum{Nunez12}].

\section{CONCLUSIONS} 

\noindent In this paper, we presented the potential of using SII in photon-counting mode for performing image synthesis, focusing on the ASTRI Mini-Array installation. We performed preliminary simulations to understand the capabilities of SI$^3$. 

\noindent The simulations show that the ASTRI Mini-Array equipped with the SI$^3$ instrument can reach performances comparable to those of larger arrays, thanks mainly to its 36 baselines. These results are encouraging and suggest that this technique can provide the capabilities needed for high-resolution imaging.

\noindent Furthermore, our simulations indicate that, when fitting 1D SII data to analytical models, it is advantageous to target stars with angular dimensions smaller than 1 mas. Conversely, for imaging purposes, it may be more profitable to observe more extended sources with substructures that produces visibility on large baselines.

\noindent Looking ahead, the development of a dedicated pipeline for image synthesis using SII data is therefore crucial. Especially considering the large amount of data to be analysed, this pipeline must be optimised as best as possible using parallelisation and HPC techniques. It will also be crucial to investigate the best algorithms to use for phase reconstruction. This will be done in the near future in parallel with the development of the instrument.

\acknowledgments 
This work is (partially) supported by ICSC – Centro Nazionale di Ricerca in High Performance Computing, Big Data and Quantum Computing, funded by European Union – NextGenerationEU.  \\
This work was conducted in the context of the ASTRI Project thanks to the support of the Italian Ministry of University and Research (MUR) as well as the Ministry for Economic Development (MISE) with funds specifically assigned to the Italian National Institute of Astrophysics (INAF).

\bibliography{report} %cambiato
\bibliographystyle{spiebib} 

\end{document}